\documentclass[twocolumn,prl,amsmath,showkeys,amssymb]{revtex4}% Physical Review Letters

\usepackage{graphicx}% Include figure files
\usepackage{dcolumn}% Align table columns on decimal point
\usepackage{bm}% bold math

\begin{document}

%\preprint{APS/123-QED}

\title{Quantum tunnelling of magnetization in Mn$_{12}$-\textit{ac} studied by $^{55}$Mn NMR }
\author{A. Morello, O. N. Bakharev, H. B. Brom, and L. J. de Jongh$^{*}$}
% \altaffiliation[Also at ]{Physics Department, XYZ University.}
%\affiliation{%
%Authors' institution and/or address\\ This line break forced with

%\affiliation{ Second institution and/or address\\
\affiliation{Kamerlingh Onnes Laboratory, Leiden Institute of Physics, Leiden University, P.O. Box 9504, 2300 RA Leiden, The Netherlands.\\
}
\date{\today}% It is always \today, today,

\begin{abstract}
We present an ultra-low temperature study (down to $T = 20$ mK) of the nuclear spin-lattice relaxation (SLR) in
the $^{55}$Mn nuclei of the molecular magnet Mn$_{12}$-\textit{ac}. The nuclear spins act as local probes for the
electronic spin fluctuations, due to thermal excitations and to tunnelling events. In the quantum regime (below $T
\approx 0.75$ K), the nuclear SLR becomes temperature-independent and is driven by fluctuations of the cluster's
electronic spin due to the quantum tunnelling of magnetization in the ground doublet. The quantitative analysis of
the nuclear SLR shows that the presence of fast-tunnelling molecules, combined with nuclear intercluster spin
diffusion, plays an important role in the relaxation process.

\end{abstract}

%;\pacs{75.45.+j}% PACS, the Physics and Astronomy
%Use showkeys class option if keyword
\maketitle

\section{Introduction}

The molecular cluster compound Mn$_{12}$-\textit{ac} [Mn$_{12}$O$_{12}$(CH$_{3}$COO)$_{16}$(H$_{2}$O)$_{4}$] was
the first single-molecule magnet to show macroscopic quantum tunnelling of magnetization (QTM) \cite{thom96}.
Since then, its magnetic properties have been studied by means of a wide variety of techniques. Nevertheless,
there are still difficulties in providing an accurate quantitative description of the tunnelling mechanism for the
giant electronic spin $S = 10$. One reason is that the real Mn$_{12}$-\textit{ac} samples contain (one or more)
minority species \cite{wern99,take02}, which differ from the majority molecules in the arrangement of the bound
H$_{2}$O and carboxylate ligands, resulting in different anisotropy barriers, easy-axes of magnetization, and
tunnelling splittings. A recent proposal also argues that a distribution of tunnelling splittings may take place
in Mn$_{12}$-\textit{ac} crystals due to the effect of dislocations \cite{chud01}. Furthermore, the study of QTM
at very low temperatures (in what we shall call the "quantum regime", i.e. where the electronic spin relaxation
rates become temperature independent) is now widely recognized to be very sensitive to the dynamics of the nuclear
spin system. So far, high-temperature NMR experiments on $^{1}$H \cite{lasc98} and $^{55}$Mn \cite{kubo02,furu01}
nuclei in Mn$_{12}$-\textit{ac} have demonstrated that the nuclear dynamics is strongly correlated with the
thermal fluctuations of the cluster's electronic spin, whereas a study of the low-T nuclear dynamics (possibly
driven by QTM) is still lacking. It is the purpose of our study to fill this experimental gap; furthermore, the
use of $^{55}$Mn nuclei as local probes for the dynamics of the electronic spins in Mn$_{12}$, opens new
possibilities for the study of QTM, since it doesn't require any macroscopic change in the electronic
magnetization of the sample.

\section{Magnetic structure and nuclear spectra}

The structure of the Mn$_{12}$-\textit{ac} molecule contains a core of four Mn$^{4+}$ ions (Mn(1)) with electronic
spin $s = 3/2$, and eight Mn$^{3+}$ ions ($s = 2$) located on an outer ring with two crystallographically
inequivalent sites (Mn(2) and Mn(3)). The superexchange interactions between ions lead to a ground-state total
spin $S = 10$ for the whole cluster. A simple spin hamiltonian for the cluster is:

\begin{eqnarray}
{\cal H} = -DS_{z}^{2} - BS_{z}^{4} - g\mu_{B}\mathbf{S}×\mathbf{B} + {\cal H}_{dip} + {\cal H}_{hyp} + {\cal H}'
\label{hamiltonian}
\end{eqnarray}

where $D = 0.55$ K is the uniaxial anisotropy parameter, $g$ = 1.94 is the gyromagnetic ratio, $S_{z}$ is the
component of the total spin $\mathbf{S}$ along the anisotropy axis and $\mathbf{B}$ is the applied magnetic
field.\\
${\cal H}' = E(S_{x}^{2} - S_{y}^{2}) + C(S_{+}^{4} + S_{-}^{4})$ is the part of the hamiltonian which does not
commute with $S_{z}$ and is therefore responsible for the tunnelling of electronic spin in zero external field.
Although the fourfold symmetry of the Mn$_{12}$ molecule should imply $E = 0$, it is now well documented that, in
a real Mn$_{12}$ sample, there are many clusters where the tetragonal symmetry is broken by the disorder in the
acetic acid molecules of crystallization \cite{corn02}. This, and the disorder in the H$_{2}$O molecules of
crystallization, is responsible for the presence of a sizable concentration (5 - 10 \%)
of clusters where the tunnelling rate can be much faster than in the "majority species".\\
${\cal H}_{dip}$ describes the dipolar interaction between electronic spins, and plays an important role in
determining the tunnelling probability for each molecule, since the dipolar field acts at low temperature as a
quasi-static bias that brings the electronic energy levels out of resonance, inhibiting the tunnelling events
\cite{prok96}.\\
The hyperfine hamiltonian ${\cal H}_{hyp}$ can be expressed as follows:

\begin{eqnarray}
{\cal H}_{hyp} = A_{1}\sum_{Mn(1)} \mathbf{I_{i1}}\cdot \mathbf{s_{i1}} + A_{2}\sum_{Mn(2)}
\mathbf{I_{i2}}\cdot \mathbf{s_{i2}}\nonumber \\
+ A_{3}\sum_{Mn(3)} \mathbf{I_{i3}}\cdot \mathbf{s_{i3}} + {\cal H}_{protons} \label{Hhyp}
\end{eqnarray}

where $A_{1}$, $A_{2}$, $A_{3}$ are the hyperfine coupling constants for the three inequivalent Mn(1), Mn(2) and
Mn(3) sites, and the sums run over the four electronic ($\mathbf{s}$) and nuclear ($\mathbf{I}$) spins at each
site. Similar but more intricate expressions would describe the coupling ${\cal H}_{protons}$ to the $^{1}$H
nuclei in the Mn$_{12}$-\textit{ac} molecule.\\
In principle the hyperfine field adds to the above mentioned bias field acting on the electronic spins due to the
intercluster dipolar interaction. If both fields would be static, tunnelling would be completely suppressed since
these effective bias fields are many orders of magnitude larger than the tunnelling splitting $\Delta_{0}$.
However, as pointed out by Prokof'ev and Stamp \cite{prok96}, by considering the dynamics of the hyperfine field
one may argue that this provides a rapidly fluctuating component that sweeps the bias field over a range much
larger than $\Delta_{0}$, thus bringing molecules in resonance for certain instants of time so that incoherent
tunnelling may occur.\\
The structure of the cluster's hamiltonian described above is reflected in the resonance spectra of the $^{55}$Mn
nuclei. Below the blocking temperature, $T_{B} \approx 3$ K, i.e. when the cluster's electronic spins are aligned
along the anisotropy $z$ axis and thermal excitations to states with $|S_{z}| < 10$ become much slower than the
experimental timescale, the hyperfine fields at the nuclear sites act effectively as strong static magnetic fields
(of order 20 - 30 T), allowing the detection of three NMR lines in zero external field. The first line (P1) is the
resonance of the nuclei belonging to the Mn$^{4+}$ ions, whereas the other two (P2 and P3) correspond to the
nuclei in the two crystallographically inequivalent Mn3+ sites. The maxima of the $^{55}$Mn resonance lines are
found at frequencies $f_{1} = 231$ MHz,  $f_{2} = 277$ MHz and  $f_{3} = 365$ MHz for P1, P2 and P3, respectively.
Since the $^{55}$Mn nuclei possess spin $I = 5/2$ and are placed in environments with symmetry lower than cubic,
each of the three NMR lines actually consists of a group of five quadrupolar-split lines. The quadrupolar
splitting $\Delta f_{Q}^{(1)} \approx 0.72$ MHz of the P1 line is much smaller than those of lines P2 and P3
($\Delta f_{Q}^{(2)} \approx 4.3$ MHz and $\Delta f_{Q}^{(3)} \approx 2.9$ MHz). Furthermore, the hyperfine field
in the Mn$^{4+}$ ions is isotropic and directed along the anisotropy axis for the $S = 10$ electronic spin of the
cluster, contrary to those in Mn$^{3+}$ ions, where there is a slight tilting \cite{kubo02}. For these reasons,
our study of the nuclear relaxation in Mn$_{12}$ was focused on the NMR signal of the line P1.

\section{Experimental}

The Mn$_{12}$ crystallites were mixed with Stycast 1266 epoxy and allowed to orient at room temperature in 9.4 T
magnetic field for one day. The oriented sample obtained in this way was mounted inside the plastic mixing chamber
of a dilution refrigerator. In the design of the refrigerator, special care was taken to ensure good
thermalization of the sample, thanks to the continuous circulation of $^{3}$He around it \cite{more03}.\\
The
nuclear spin-lattice relaxation was investigated by means of the pulse NMR technique, monitoring the recovery of
the nuclear magnetization $M(t)$ after an inversion pulse. Typically, the duration of a $\pi /2$ pulse was 10 ms.
Since the $^{55}$Mn nuclei have spin $I = 5/2$, the magnetization recovery is not described by a single
exponential. The prediction for the recovery after inversion of the central of the five quadrupolar split lines,
in the limit where the quadrupolar splitting is much smaller than the Zeeman splitting, has been obtained as
\cite{sute98}:

\begin{eqnarray}
    \frac{M(t)}{M(\infty)} = 1 - [ \frac{100}{63} \exp(-30 W t) + \frac{16}{45} \exp(-12 W t) \nonumber \\
    + \frac{2}{35} \exp(-2 W t)]
\label{recovery}
\end{eqnarray}
where $W$ is the nuclear spin-lattice relaxation rate. Eq. \ref{recovery} is found to fit very well the whole
recovery curve, leaving only $W$ as (time-related) fitting parameter.

\section{Results and discussion}

The nuclear spin-lattice relaxation rate for the $^{55}$Mn nuclei of Mn$_{12}$ in the temperature range 1 - 3 K,
has been previously studied by Furukawa et al. \cite{furu01}. In that work, $T_{1}^{-1}$ ($= 2W$) was obtained
from the magnetization recovery after saturation of the central line, fitting only the initial part of $M(t)$. Our
results in the range 1 - 2 K, shown in Fig. 1, are in good agreement with those of Ref. \cite{furu01}, despite the
use of different pulse sequences and fitting functions for the recovery curves. As mentioned in the introduction,
the $^{55}$Mn nuclei are used as local probes for the fluctuations of the cluster's electronic spin. From the
temperature dependence of the nuclear relaxation rate $W(T)$ above 1 K, it was deduced \cite{furu01} that the
electronic fluctuations arise from thermal excitations from the ground levels $|S_{z}| = 10$ to the nearest
excited levels $|S_{z}| = 9$ inside each energy potential well, i.e. without crossing the anisotropy barrier. Such
excitations produce a fluctuating transverse magnetic field $\langle h_{\perp} \rangle$ at the nuclear site. Its
time-correlation function can be written as $\langle h_{\perp}(0) h_{\perp}(t) \rangle = \langle h_{\perp}^{2}
\rangle \exp (-t/\tau) $ where $\tau$ is the typical timescale for the fluctuations. Above 1 K, $\tau$ is
determined by the timescale $\tau_{s-ph}$ of the spin-phonon excitations. Since in the investigated temperature
range (up to 2 K) only the thermal excitations to the nearest excited level need be considered, then $\tau_{s-ph}
\sim \tau_{0} \exp[(E_{9} - E_{10}) / k_{B}T]$, where $E_{9} - E_{10}$ is the energy difference between the ground
and the first excited state \cite{furu01, leue00}. In this limit, the nuclear relaxation rate can be expressed as:

\begin{equation}
W \approx \frac{\gamma_{n}^{2}}{4} \langle h_{\perp}^{2} \rangle \frac{\tau_{s-ph}}{1 +
\omega_{N}^{2}\tau_{s-ph}^{2}}
\label{W}
\end{equation}

where $\gamma_{N} = 10.57$ MHz/T is the nuclear gyromagnetic ratio, and $\omega_{N} / 2\pi = 231$ MHz is the
Larmor frequency. The solid line in Fig. 1 was obtained fixing $E_{9} - E_{10} = 14.58$ K, as calculated from the
spin hamiltonian (\ref{hamiltonian}) using the experimental values $D = 0.55$ K and $B = 1.2 \times 10^{-3}$ K
(see [7] and references therein), and obtaining $\langle h_{\perp}^{2} \rangle / \tau_{0} \approx 2.8 \times
10^{8}$ T$^{2}$/s. The fit could be slightly improved by leaving $E_{9} - E_{10}$ as free parameter, yielding
$E_{9} - E_{10}\approx 12.1$ K and $\langle h_{\perp}^{2} \rangle / \tau_{0} \approx 4.5 \times 10^{7}$ T$^{2}$/s.
This demonstrates that (\ref{W}) indeed correctly describes the experimental data in the thermally activated
regime
for the electronic spin fluctuations.\\

\begin{figure}
\includegraphics[width=8cm]{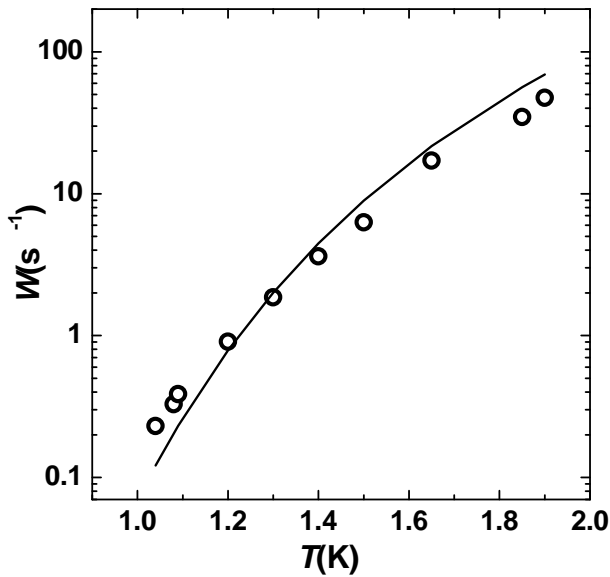}
\caption{\label{fig1} Nuclear SLR rate above 1 K. Solid line: fit to Eq. (4), showing that the nuclear relaxation
is produced by thermal fluctuations of the electronic spin.}
\end{figure}

Given the temperature dependence of $\tau_{s-ph}$, the above mechanism would lead to astronomically long nuclear
relaxation at millikelvin temperatures. The main goal of our research was to find out whether the nuclear spins
would still be able to relax at very low temperature, thanks to fluctuations due to incoherent quantum tunnelling
of the cluster's electronic spin within the magnetic ground ($S_{z} = \pm 10$) doublet. The possibility for such
tunnelling to occur even for extremely small values of the tunnelling splitting, was argued by Prokof'ev and Stamp
\cite{prok96}, on basis of a formalism invoking the dynamics of the hyperfine interactions due to nuclear spin
diffusion. Fig. 2 shows the nuclear relaxation rate as measured down to $T = 20$ mK. Obviously, another relaxation
mechanism is present, since $W(T)$ becomes temperature independent below $T \approx 0.75$ K. It is very tempting
to ascribe such a behavior to QTM, also because at a similar temperature the hysteresis loops for the cluster's
magnetization were found to become temperature independent \cite{chio00}. In fact, an equivalent phenomenon has
very recently been observed for the proton NMR relaxation in Fe$_{8}$ \cite{ueda02}, and also in that case the
plateau in $W(T)$ appears at the same temperature where the magnetization loops become temperature independent.
Additional evidence supporting the idea of tunnelling-driven nuclear relaxation is provided by the dependence of
$W$ on external magnetic field $B_{z}$ applied along the anisotropy axis of the clusters. A comprehensive study of
$W(B_{z})$ at various
temperatures is in preparation and will be published elsewhere.\\
The very fast nuclear relaxation rate $W \approx 0.03$ s$^{-1}$ in the quantum regime is, at a first glance, quite
astonishing. One could think of explaining it in terms of magnetic fluctuations directly felt by the nuclear spins
as a consequence of a tunnelling event, thus using (\ref{W}) and replacing $\langle h_{\perp}^{2} \rangle$ and
$\tau_{s-ph}$ by the amplitude and timescale for the fluctuations arising from the tunnelling of the cluster's
electronic spin, but the result would be several orders of magnitude lower than the observed value. One
possibility is to invoke the role of fast-relaxing molecules, which have a much higher tunnelling rate. However we
can demonstrate that the signal we measure comes from the nuclei in the Mn$^{4+}$ ions of all Mn$_{12}$ clusters
in the sample, i.e. not only those belonging to fast-relaxing molecules. When applying an external magnetic field
$B_{z}$ along the anisotropy axis of the clusters, this can add or subtract to the hyperfine field at the nuclear
site, depending on whether the electronic spin is parallel or antiparallel to the applied field. Therefore, some
nuclei will have their Larmor frequencies shifted up (when $B_{z}$ is parallel to the electronic spin $S$), the
others down ($B_{z}$ antiparallel to $S$). By measuring the nuclear spin-echo signal intensity at the two possible
Larmor frequencies, we can check the magnetization state of the sample (see also Ref. \cite{kubo02}), and we
observe that, starting from a fully magnetized sample, its electronic magnetization (as seen from the nuclei) does
not substantially relax even after one week of measurements. This implies that the nuclei we observe belong to all
clusters, i.e. also those for which the tunnelling rates are extremely slow.\\

\begin{figure}
\includegraphics[width=8cm]{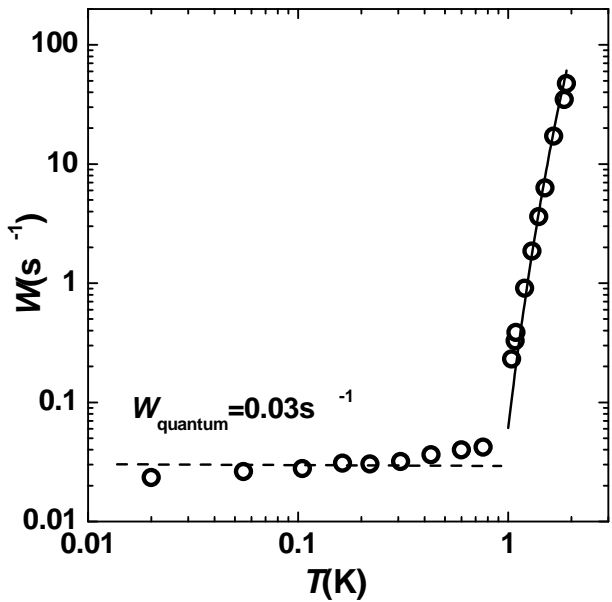}
\caption{\label{fig2} Below $T \approx 0.75$ K the nuclear SLR becomes temperature independent and saturates at a
value $W_{quantum} \approx 0.03$ s$^{-1}$. This reveals the presence of (temperature independent) tunnelling
fluctuations.}
\end{figure}

In view of the important role of nuclear spin diffusion in the model of Prokof'ev and Stamp, where it provides the
dynamics of the hyperfine field acting on the electron spin, it seems likely that a correct quantitative
interpretation of the nuclear relaxation rate should involve the role of nuclear spin diffusion as well. We have
therefore measured the transverse nuclear spin-spin relaxation rate $T_{2}^{-1}$, shown in  Fig. 3. Its
low-temperature value $T_{2}^{-1} \approx 100$ s$^{-1}$, agrees with the nuclear spin diffusion rate that we can
calculate by taking into account the flip-flop term in the dipolar interaction between nuclei of Mn$^{4+}$ ions
belonging to all neighboring clusters. This means that the intercluster spin diffusion is an effective mechanism
to transport nuclear polarization across the sample, on a timescale much shorter than the observed spin-lattice
relaxation. The value of $1 / (2WT_{2}) \sim 10^{3}$ means that about $10^{3}$ nuclear flip-flop events can take
place during the nuclear spin-lattice relaxation time. Accordingly, the energy of the whole system of $^{55}$Mn
nuclei can be effectively transported by nuclear spin diffusion to the location of the fast-tunnelling electronic
spins, which provide the necessary relaxation channel by incoherent tunnelling events.

\begin{figure}
\includegraphics[width=8cm,height=8cm]{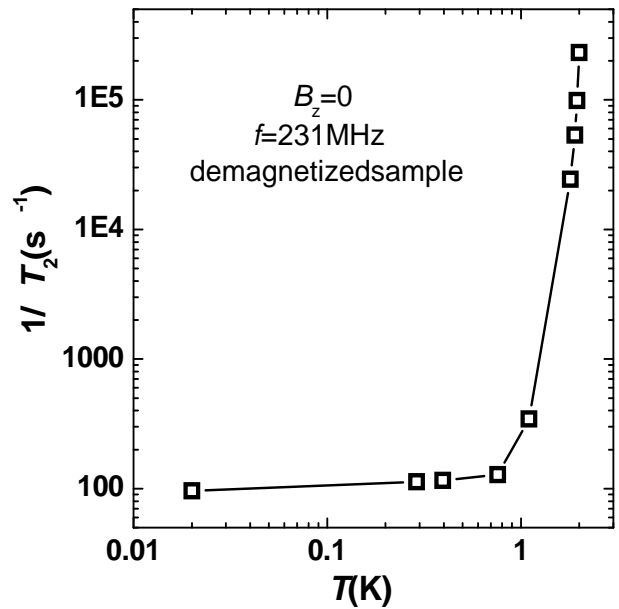}
\caption{\label{fig3} Transversal nuclear spin-spin relaxation rate: the low-$T$ value agrees with the calculated
intercluster spin diffusion rate.}
\end{figure}

\section{Conclusions}

We have shown that the nuclear spins can be used as effective local probes for the detection of tunnelling
fluctuations in the electronic spin of Mn$_{12}$-\textit{ac}. In particular, the nuclear relaxation rate $W$
becomes temperature independent below $T \approx 0.75$ K. The quantitative analysis of $W$ suggests that the
presence of a small concentration of fast-tunnelling electronic spins, combined with nuclear intercluster spin
diffusion, provides the mechanism responsible for the observed nuclear spin-lattice relaxation rate.

\section{Acknowledgements}

We thank Roberta Sessoli and Andrea Caneschi for providing the Mn$_{12}$-\textit{ac} sample and for fruitful
discussions. We also acknowledge helpful and stimulating discussions with Boris Fine and Philip Stamp. This work
is part of the research program of the "Stichting voor Fundamenteel Onderzoek der Materie" (FOM).\\

\noindent $^*$ Corresponding author:

L. J. de Jongh, e-mail: dejongh@phys.leidenuniv.nl

\bibliographystyle{unsrt}
 \bibliography{icmm02condmat}

\end{document}